\documentclass[doublecol]{epl2}
\usepackage{amssymb,amsmath,graphicx,setspace,color,rotating,subfigure,url,times}
\title{Emergence of long memory in stock volatility from a modified Mike-Farmer model}

\author{Gao-Feng Gu\inst{1,2,3} \and Wei-Xing Zhou\inst{1,2,3,4,5}\footnote{e-mail: wxzhou@ecust.edu.cn%
}} \shortauthor{G.-F. Gu \etal}

\institute{
  \inst{1} School of Business, East China University of Science and Technology, Shanghai 200237, China\\
  \inst{2} School of Science, East China University of Science and Technology, Shanghai 200237, China\\
  \inst{3} Research Center for Econophysics, East China University of Science and Technology, Shanghai 200237, China\\
  \inst{4} Research Center on Fictitious Economics \& Data Science, Chinese Academy of Sciences, Beijing 100080, China\\
  \inst{5} Engineering Research Center of Process Systems Engineering (Ministry of Education), East China University of Science and Technology, Shanghai 200237, China%
}


\abstract{The Mike-Farmer (MF) model was constructed empirically
based on the continuous double auction mechanism in an order-driven
market, which can successfully reproduce the cubic law of returns
and the diffusive behavior of stock prices at the transaction level.
However, the volatility (defined by absolute return) in the MF model
does not show sound long memory. We propose a modified version of
the MF model by including a new ingredient, that is, long memory in
the aggressiveness (quantified by the relative prices) of incoming
orders, which is an important stylized fact identified by analyzing
the order flows of 23 liquid Chinese stocks. Long memory emerges in
the volatility synthesized from the modified MF model with the DFA
scaling exponent close to 0.76, and the cubic law of returns and the
diffusive behavior of prices are also produced at the same time. We
also find that the long memory of order signs has no impact on the
long memory property of volatility, and the memory effect of order
aggressiveness has little impact on the diffusiveness of stock
prices.}

\begin{document}

\maketitle

\section{Introduction}

The continuous double auction mechanism is adopted in the electronic
trading systems in many stock markets worldwide. In particular, most
emerging stock markets are order-driven markets. In a pure
order-driven market, there are no market makers or specialists, and
market participants submit and cancel orders, which may result in
transactions based on price-time priority. Different from
quote-driven markets where market makers are liquidity providers,
the same trader in an order-driven market can act as either a
liquidity taker or a liquidity provider depending on the
aggressiveness of her submitted orders. The behaviors of market
makers are very complicated, since they have the obligation to
maintain the liquidity of stocks and in the meanwhile want to
maximize their profits. It is thus natural to argue that it is
easier to construct microscopic models for order-driven markets than
for quote-driven markets in order to understand the macroscopic
regularities of stock markets from a microscopic angle of view.

Indeed, a lot of efforts have been made to construct order-driven
models \cite{Slanina-2008-EPJB}, which can be dated back to the
1960's \cite{Stigler-1964-JB}. In order to check if the model
captures some basic aspects of the underlying mechanisms governing
the evolution of stock prices, one usually investigates the
statistical properties of the mock stocks, such as the distribution
and autocorrelation of returns and the long memory in volatility.
Deviations from these well-established stylized facts allow us to
improve the models and gain a better understanding of the underlying
microscopic mechanisms. For instance, the DFA scaling exponent of
price fluctuations is found to be significantly less than the
empirical value in the Bak-Paczuski-Shubik model
\cite{Bak-Paczuski-Shubik-1997-PA} and in the Maslov model
\cite{Maslov-2000-PA}, leading to new order-driven models
\cite{Willmann-Schutz-Challet-2002-PA,Preis-Golke-Paul-Schneider-2006-EPL,Preis-Golke-Paul-Schneider-2007-PRE,Svorencik-Slanina-2007-EPJB}.

Recently, Mike and Farmer have proposed an empirical behavioral
model, which is based on the statistical properties of order
placement and cancelation extracted from ultrahigh-frequency stock
data \cite{Mike-Farmer-2008-JEDC}. To the best of our knowledge, the
Mike-Farmer model (or MF model for short) is the only
{\em{empirical}} model, which outperforms other order-driven models
and is adaptive for further improvement. The MF model can reproduce
several important stylized facts: The returns are distributed
according to the cubic law, the DFA scaling exponent of returns is
close to 0.5, and the spreads and lifetimes of orders have power-law
tails. However, the DFA scaling exponent of the volatility is also
found to be $H_v\approx0.6$, which is much less than the empirical
value of $H_v\approx0.8$ \cite{Mike-Farmer-2008-JEDC}. In this work,
we propose a modified version of the MF model, which is able to
produce very realistic strong persistence in the volatility without
destruction of other stylized facts.

The volatility clustering phenomenon, as well as other important
stylized facts, can be observed in many other microscopic market
models. In the econophysics literature, physicists model stock
markets as a complex system with interacting agents and different
physics scenarios lead to different types of models
\cite{Zhou-2007}, such as percolation models
\cite{Cont-Bouchaud-2000-MeD,Stauffer-1998-AP,Stauffer-Penna-1998-PA,Eguiluz-Zimmermann-2000-PRL,DHulst-Rodgers-2000-IJTAF,Xie-Wang-Quan-Yang-Hui-2002-PRE},
spin models
\cite{Foellmer-1974-JMathE,Chowdhury-Stauffer-1999-EPJB,Iori-1999-IJMPC,Kaizoji-2000-PA,Bornholdt-2001-IJMPC,Zhou-Sornette-2007-EPJB},
minority games
\cite{Arthur-1994-AER,Challet-Zhang-1997-PA,Challet-Marsili-Zhang-2000-PA,Jefferies-Hart-Hui-Johnson-2001-EPJB,Challet-Marsili-Zhang-2001-QF,Challet-Marsili-Zhang-2001a-PA,Challet-Marsili-Zhang-2001b-PA,Challet-Marsili-Zhang-2005},
majority games
\cite{Marsili-2001-PA,Kozlowski-Marsili-2003-JPA,Gou-2005-PA}, and
the \$-game \cite{Andersen-Sornette-2003-EPJB}, to list a few. There
is also a long list of stock market models in the economics
literature \cite{LeBaron-2000-JEDC}. In contrast with models where
the agents (or traders) are homogenous, most of economic models
assume that the traders have bounded rationality and heterogeneous
beliefs \cite{Brock-Hommes-1997-Em,Brock-Hommes-1998-JEDC}. Traders
can thus be classified into two types: fundamentalists and
chartists. The fundamentalists believe that the asset price is
solely determined by economic fundamentals and they buy (or sell)
when the price is lower (or higher) than the fundamental price. On
the contrary, chartists are trend followers and try to predict
future price movement according to diverse techniques. Many
theoretical and computational oriented models have been proposed
\cite{Lux-Marchesi-1999-Nature,Gaunersdorfer-2000-JEDC,LeBaron-2001-MeD,Hommes-2001-QF,Chiarella-He-2001-QF,Hommes-2002-PNAS,Brock-Hommes-Wagener-2005-JMathE,Boswijk-Hommes-Manzan-2007-JEDC,Amilon-2008-JEF}.

\section{Mike-Farmer model and its modification}
\label{s2:MFM}

The MF model contains two main parts, order placement and
cancelation. In order to submit an order, one needs to decide its
direction (buy or sell), price and size. In the MF model, the size
of any order is fixed to one. The sign of orders presents strong
long memory, with $H_s\approx0.8$ \cite{Lillo-Farmer-2004-SNDE}.
Therefore, order signs can be generated from fractional Brownian
motions with DFA scaling exponent $H_s$. The price of an incoming
order can be characterized by the relative price $x$, which is the
logarithmic distance of the order price to the same best price:
\begin{equation}
x(t) = \left\{
 \begin{array}{ccl}
 \ln \pi(t)  - \ln \pi_b(t-1), && ~~{\rm{buy~orders}} \\
 \ln \pi_a(t-1) - \ln \pi(t), && ~~{\rm{sell~orders}}
 \end{array}
 \right.,
\label{Eq:x}
\end{equation}
where $\pi(t)$ is the order price at time $t$, and $\pi_b(t-1)$ and
$\pi_a(t-1)$ are the best bid and best ask at time $t-1$,
respectively. The relative prices in the MF model are generated from
a Student distribution whose degrees of freedom $\alpha_x$ and
scaling parameter $\sigma_x$ are determined empirically using real
stock data. Mike and Farmer also proposed a model for order
cancelation combining three factors: the position of an order in the
order book, the imbalance of buy and sell orders in the book, and
the total number of orders in the book.

With these findings in hand, our simulations of the MF model can be
described as follows. Before the evolution of prices, we generate an
array of relative prices $\{x(t):t=1,2,\cdots,T\}$, drawn from the
Student distribution with $\alpha_x = 1.3$ and $\sigma_x = 0.0024$,
and an array of signs $\{s(t):t=1,2,\cdots,T\}$ according to a
fractional Brownian motion with $H_s = 0.75$. At each simulation
step $t$, an order is generated, whose relative price and direction
are $x(t)$ and $s(t)$, respectively. If $x(t)$ is not less than the
spread, the order is an effective market order, resulting in an
immediate execution with a limit order waiting at the opposite best
price. Otherwise, the incoming order is an effective limit order,
which is stored in the queue of the limit order book. Then we scan
the standing orders to check if any of them can be canceled,
following exactly the same process in the MF model. We simulate $T=2
\times 10^5$ steps in each round. The stock prices are recorded and
we analyze the last $4 \times 10^4$ returns in each round.

\begin{figure}[htb]
\centering
\includegraphics[width=7cm]{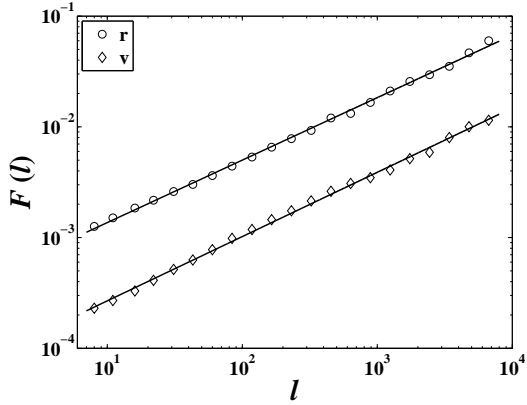}
\caption{\label{Fig:MF:DFA:RV} Detrended fluctuation function
$F(\ell)$ as a function of time lag $\ell$ for the returns and the
volatility, respectively. The solid lines are the linear
least-squares fits to the data and $H_r = 0.55 \pm 0.01$ for returns
and $H_v = 0.58 \pm 0.01$ for volatility. The plot for volatility
has been shifted vertically for clarity.}
\end{figure}

The distribution of returns in the MF model has been studied in
detail and we reproduced the cubic law \cite{Gu-Zhou-2009-EPJB}. We
now perform a detrended fluctuation analysis (DFA)
\cite{Peng-Buldyrev-Havlin-Simons-Stanley-Goldberger-1994-PRE,Kantelhardt-Bunde-Rego-Havlin-Bunde-2001-PA}
on the return $r$ and the volatility $v = |r|$ to estimate the DFA
scaling exponents. The results are shown in
Fig.~\ref{Fig:MF:DFA:RV}. Excellent power-law dependence of the
detrended fluctuation function $F(\ell)$ with respect to the
timescale $\ell$ is observed for the two quantities in the scaling
range $8 \leqslant \ell < 7000$. The DFA scaling exponents are $H_r
= 0.55$ for the returns and $H_v = 0.58$ for the volatility,
respectively. These indexes are merely a little greater than 0.5,
which means that there is no long memory or very weak memory in the
returns and the volatility. To obtain a solid picture, we repeated
the simulations of the MF model 20 times and performed DFA on the
returns and the volatility. We find that $H_r$ varies in the range
$[0.54, 0.58]$ with the average $\overline{H}_r = 0.57 \pm 0.01$ for
the returns, and $H_v$ varies in the range $[0.56, 0.62]$ with the
average $\overline{H}_v = 0.59 \pm 0.01$ for the volatility. This
analysis confirms the results of Mike and Farmer
\cite{Mike-Farmer-2008-JEDC}. It is well accepted in mainstream
Finance that there is no memory in returns
\cite{Bouchaud-Gefen-Potters-Wyart-2004-QF}, consistent with the
weak-form market efficiency hypothesis, while the volatility possess
strong persistence with the DFA scaling exponent much greater than
0.5 \cite{Mantegna-Stanley-2000}. Therefore, the MF model captures
the stylized fact that $H_r$ of returns is close to 0.5, but fails
to reproduce strong memory effect in the volatility. Obviously,
certain important feature is missing in the original MF model, which
calls for a further scrutiny of the real stock data and a
modification of the model.

In financial markets, it is impossible for a trader to collect and
digest all information that is available publicly, and it is not
free to collect and process diverse information from different
sources. Due to the limited processing power of human brains and
finite amount of money, it is not irrational for traders to mimic
the trading behaviors of others, which may lead to positive
feedbacks and herding behaviors in an intermittent fashion. In other
words, most traders in financial markets play a majority game. They
are more willing to buy when the price rises and to sell when the
price falls. This scenario is known as the information cascading
mechanism \cite{Alevy-Haigh-List-2007-JF} and it is well documented
that imitation and herding cause the emergence of volatility
clustering and long memory. A comprehensive taxonomy of herd
behavior was synthesized by Hirshleifer and Teoh
\cite{Hirshleifer-Teoh-2003-EFM}. We also refer to an excellent book
of Lyons for a modern treatment \cite{Lyons-2001}. Following this
line, a trader is very possible to submit an order that is
``similar'' to its preceding limit orders. In addition, the long
memory in the order flow is well-known as ``diagonal effect''
\cite{Biais-Hillion-Spatt-1995-JF}. Other than herding, there are at
least two alternative hypotheses for the origin of long memory in
the order flow: order splitting and traders reacting similarly to
the same signal \cite{Biais-Hillion-Spatt-1995-JF}. Since an order
is fully determined by its direction (order sign), aggressiveness
(order price) and size, we expect that these variables might also
have strong memory. In the MF model, the directions of incoming
orders are modeled by fractional Brownian motions with $H_s\gg 0.5$,
while the order size is fixed. It is thus worthwhile to check if the
order aggressiveness characterized by relative prices has long
memory using real ultrahigh-frequency stock data, and if the long
memory in the order aggressiveness, if any, can cause the emergence
of long memory in the volatility.

\begin{figure}[htb]
\centering
\includegraphics[width=7cm]{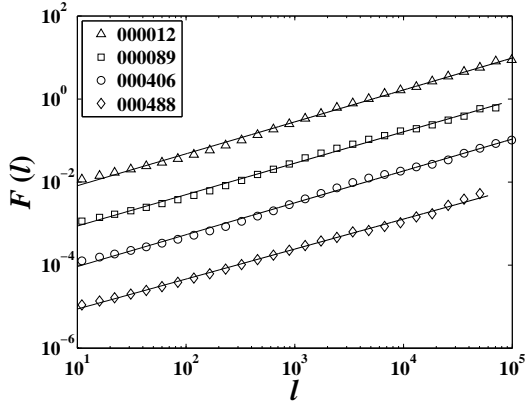}
\caption{\label{Fig:DFA:x} Dependence of the detrended fluctuation
function $F(\ell)$ with respect to the timescale $\ell$ for four
stocks, whose stock codes are 000012, 000089, 000406 and 000488. The
solid lines are the linear least-squares fits to the data and $H_x =
0.77 \pm 0.01$, $0.76 \pm 0.01$, $0.77 \pm 0.01$, and $0.72 \pm
0.01$, respectively. The plots for stocks 000089, 000406 and 000488
have been shifted vertically for clarity.}
\end{figure}

In order to study the memory effect of order aggressiveness, we
utilize a nice database of 23 liquid stocks listed on the Shenzhen
Stock Exchange in the whole year 2003 \cite{Gu-Chen-Zhou-2008a-PA}.
The database contains detailed information of the incoming order
flow, such as order direction and size, limit price, time, best bid,
best ask, transaction volume, and so on. We focus on the relative
prices of orders submitted during the continuous double auction.
Figure~\ref{Fig:DFA:x} illustrates the dependence of the detrended
fluctuation functions $F(\ell)$ with respect to the timescale $\ell$
for four randomly chosen stocks. Sound power-law scaling relations
are observed in the scaling ranges spanning four orders of
magnitude. The DFA scaling exponents of the relative prices for the
four stocks are estimated to be $H_x = 0.77 \pm 0.01$ in the scaling
range $10 \leqslant \ell < 10^5$, $0.76 \pm 0.01$ in the scaling
range $10 \leqslant \ell < 7\times10^4$, $0.77 \pm 0.01$ in the
scaling range $10 \leqslant \ell < 10^5$, and $0.72 \pm 0.01$ in the
scaling range $10 \leqslant \ell < 5\times10^4$, respectively. The
DFA results for other stocks are quite similar. We find that $H_x$
varies in the range $[0.72, 0.87]$ with an average $\overline{H}_x =
0.78 \pm 0.03$. It is evident that the relative price $x$ is
super-diffusive and possesses long-term dependence.

It is noteworthy to point out that the long memory temporal
structure in the relative prices was also observed in the London
Stock Exchange. Zovko and Farmer studied the autocorrelation
function of relative prices for buy orders and sell orders of 50
stocks traded on the London Stock Exchange
\cite{Zovko-Farmer-2002-QF}. They found that the autocorrelation
function decays as a power law with exponent $\gamma=0.41\pm0.07$.
It follows immediately that $H_x=1-\gamma/2=0.80\pm0.04$
\cite{Samorodnisky-Taqqu-1994}. We also performed detrended
fluctuation analysis of the relative prices for buy orders and sell
orders of the four stocks analyzed in Fig.~\ref{Fig:DFA:x}. The
exponents are $0.75 \pm 0.01$, $0.81 \pm 0.01$, $0.77 \pm 0.01$ and
$0.70 \pm 0.01$ for buy orders and $0.77 \pm 0.01$, $0.75 \pm 0.01$,
$0.75 \pm 0.01$ and $0.71 \pm 0.01$ for sell orders. There is no
significant difference in the memory properties if one considers
relative prices of orders on the same side of the book.

Based on the above empirical finding that the relative prices have
long memory, we can introduce a new ingredient in the MF model. The
modified MF model inherits all the ingredients of the MF model
except that the relative prices are generated from a Student
distribution with long memory. This can be done as follows. We
generate an array of relative prices $\{x_0(t):t=1,2,\cdots,T\}$
from a Student distribution. Then we simulate a fractional Brownian
motion with $H_x=0.8$ and record its differences as
$\{y(t):t=1,2,\cdots,T\}$. The sequence $\{x_0(t):t=1,2,\cdots,T\}$
is rearranged such that the rearranged series
$\{x(t):t=1,2,\cdots,T\}$ has the same rank ordering as
$\{y(t):t=1,2,\cdots,T\}$, that is, $x(t)$ should rank $n$ in
sequence $\{x(t):t=1,2,\cdots,T\}$ if and only if $y(t)$ ranks $n$
in the $\{y(t):t=1,2,\cdots,T\}$ sequence
\cite{Bogachev-Eichner-Bunde-2007-PRL,Zhou-2008-PRE}. It is obvious
that $x(t)$ still obeys the same Student distribution. A detrended
fluctuation analysis of $x(t)$ shows that its DFA scaling exponent
is very close to $H_x=0.8$. This sequence of $x(t)$ is used as the
relative prices in our modified MF model.

\section{Numerical results}

\begin{figure}[htb]
\centering
\includegraphics[width=7cm]{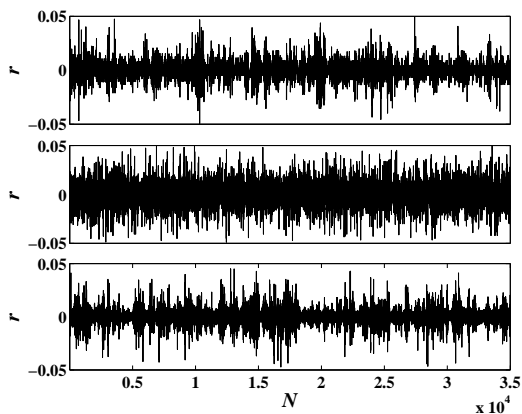}
\caption{\label{Fig:3Returns} Comparison of typical return time
series from a real Chinese stock 000012 (upper panel), the original
MF model (middle panel), and the modified MF model (lower panel).}
\end{figure}

Based on the modified MF model discussed above, we first generate
the relative prices $x$ from the Student distribution with
parameters $\alpha_x = 1.3$ and $\sigma_x = 0.0024$. Then we add
long memory to the time series, using $H_x \approx 0.8$. In each
round, we simulate the modified MF model $2\times10^5$ steps with
the same parameters $H_s = 0.75$, $A = 1.12$ and $B = 0.2$ and
record the return time series with the length near $4 \times 10^4$
after removing the transient period. In Fig.~\ref{Fig:3Returns}, we
illustrate a typical segment of the simulated returns from the
modified MF model, which is compared with the return time series of
a real Chinese stock (code 000012) and the original MF model. It is
evident that the return time series of the modified MF model
exhibits clear clustering resembling the clustering phenomenon in
real data, whereas the simulated returns from the original MF model
do not show clear clustering feature. This already indicates
qualitatively that the volatility of the modified MF model has
stronger long-term memory than that of the original MF model.

\begin{figure}[htb]
\centering
\includegraphics[width=7cm]{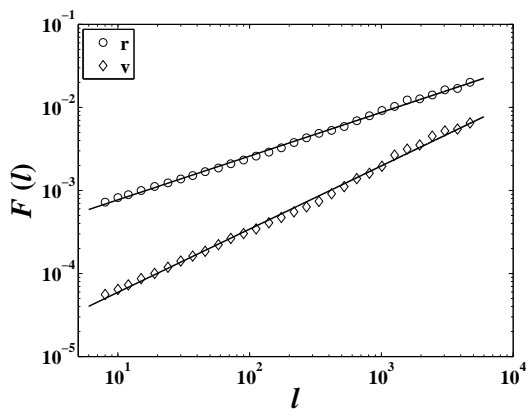}
\caption{\label{Fig:MMF:DFA} Detrended fluctuation analysis of the
returns $r$ and the volatility $v$ generated according to the
modified MF model. The solid lines are the linear least-squares fits
to the data and $H_r = 0.53 \pm 0.01$ for the returns and $H_v =
0.76 \pm 0.01$ for the volatility. The plot for volatility has been
shifted vertically for clarity.}
\end{figure}

To quantify the strength of the memory effect in the simulated
volatility, we have performed the detrended fluctuation analysis.
Figure~\ref{Fig:MMF:DFA} shows the dependence of the detrended
fluctuation $F(\ell)$ as a function of the timescale $\ell$ in
log-log coordinates. We find that $F(\ell)$ scales as a power law
against $\ell$ with the scaling range spanning about three orders of
magnitude. We obtain $H_v = 0.76 \pm 0.01$ in the scaling range $8
\leqslant \ell < 4500$, which is in excellent agreement with
empirical results. We also performed a detrended fluctuation
analysis on the returns. The results are also presented in
Fig.~\ref{Fig:MMF:DFA}. We find that $H_r = 0.53 \pm 0.01$ in the
scaling range $8 \leqslant \ell < 4500$, consistent with empirical
results. Comparing with Fig.~\ref{Fig:MF:DFA:RV}, we conclude that
the value of $H_x$ has little impact on $H_r$. We repeated this
process for 20 times and the results are very similar. The exponent
$H_v$ varies in the range $[0.74, 0.77]$ with an average
$\overline{H}_v = 0.76 \pm 0.01$, while $H_r$ ranges in $[0.53,
0.55]$ with an average $\overline{H}_r = 0.54 \pm 0.01$.

In order to further inspect the quantitative relation between $H_x$
and $H_v$, more simulations with different values of $H_x$ have been
performed. For each fixed $H_x$, repeated simulations do not show
much fluctuation in $H_v$. The results are shown in Table
\ref{TB:HvHr:Hx}. It is found that $H_v$ is not identical to $H_x$.
However, $H_v$ increases with $H_x$. Table \ref{TB:HvHr:Hx} also
confirms that $H_r$ is close to 0.5 and independent of $H_x$. The
relation between volatility clustering and relative prices has been
detected and investigated for stocks on the London Stock Exchange
\cite{Zovko-Farmer-2002-QF}.

\begin{table}[htb]
\centering
\caption{\label{TB:HvHr:Hx}Dependence of $H_v$ and $H_r$ on $H_x$. %
   For each value of $H_x$, ten repeated simulations are conducted.
   The scaling range is $8 \leqslant \ell < 4500$.
   The numbers in the parentheses are the standard deviations divided by 100. }%
\medskip
\begin{tabular}{ccccccc}
\hline\hline
 $H_x$  &   0.50    &   0.60    &    0.70   &    0.80   &   0.90    \\\hline%
 $H_v$  & $0.57(1)$ & $0.61(1)$ & $0.67(1)$ & $0.76(1)$ & $0.81(2)$ \\%
 $H_r$  & $0.55(1)$ & $0.55(1)$ & $0.54(1)$ & $0.54(1)$ & $0.54(1)$ \\%
 \hline\hline
\end{tabular}
\end{table}

\begin{figure}[htb]
\centering
\includegraphics[width=7cm]{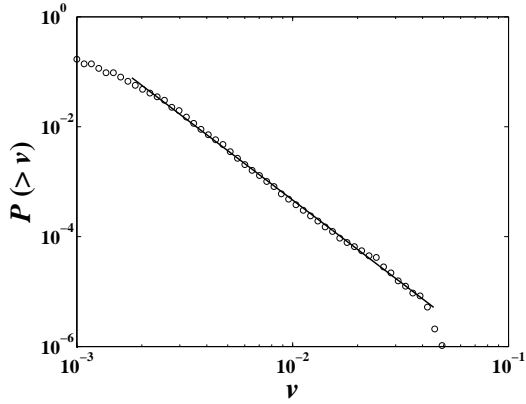}
\caption{\label{Fig:MMF:CDF} Empirical complementary cumulative
distribution $P(>v)$ of the volatility generated according to the
modified MF model in double logarithmic coordinates. The solid line
is the best power-law fit to the data with the tail index $\beta =
2.99 \pm 0.02$.}
\end{figure}

Figure \ref{Fig:MMF:CDF} shows the empirical complementary
cumulative distribution $P(>v)$ of the volatility generated
according to the modified MF model. We find that the volatility has
a power-law tail
\begin{equation}
 P(> v) \sim v^{-\beta},
 \label{Eq:MMF:P:v}
\end{equation}
where $\beta$ is the tail index. Using the least-squares fitting
method, we obtain that $\beta = 2.99 \pm 0.02$, identical to 3. In
other words, the volatility obeys the well-known cubic law
\cite{Gopikrishnan-Meyer-Amaral-Stanley-1998-EPJB}, which is
captured by the original MF model
\cite{Mike-Farmer-2008-JEDC,Gu-Zhou-2009-EPJB}.

Additional numerical experiments show that the cancelation process
in the modified MF model is not the only one to reproduce the main
stylized facts. The modified MF model with a Poissonian cancelation
process gives $H_r = 0.51 \pm 0.01$, $H_v = 0.81 \pm 0.01$, and
$\beta = 3.19 \pm 0.03$.

\begin{figure}[h!]
\centering
\includegraphics[width=7cm]{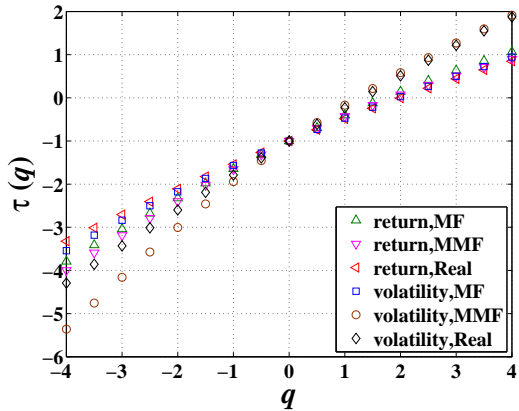}
\caption{\label{Fig:Tau} Multifractal detrended fluctuation analysis
of the returns $r$ and the volatility $v$ generated according to the
MF model and the modified MF model with comparison to the
multifractal nature of the real data. }
\end{figure}

Beside efficiency and long memory of the volatility and the cubic
law of the return, the price dynamics is characterized by
multifractality \cite{Mantegna-Stanley-2000}. We adopted the
multifractal detrended fluctuation analysis
\cite{Kantelhardt-Zschiegner-Bunde-Havlin-Bunde-Stanley-2002-PA} to
investigate the return and volatility time series generated from the
MF model, the modified MF model, and the real data as well for
comparison. For a given time series, the $q$-th order detrended
fluctuation function $F_q(s)$ scales as a power law
\begin{equation}
 F_q(s) \sim s^{H(q)}~
 \label{Eq:Fq:s}
\end{equation}
and the mass exponent $\tau(q)$ in the standard textbook structure
function formalism is
\cite{Kantelhardt-Zschiegner-Bunde-Havlin-Bunde-Stanley-2002-PA}
\begin{equation}
 \tau(q) =qH(q)-1~.
 \label{Eq:Tau:Hq}
\end{equation}
Note that $H(q=2)$ is the DFA scaling exponent characterizing the
long memory property of the time series. The mass exponent $\tau(q)$
of each financial variable is plotted in Fig.~\ref{Fig:Tau} as a
function of $q$. When $q=0$, $\tau(0)=-1$ for each case, as
predicted by Eq.~(\ref{Eq:Tau:Hq}). It is evident that all $\tau(q)$
functions are nonlinear with respect to $q$, which confirms the
multifractal nature of return and volatility in both models and in
real data. When $q<0$, both models deviate remarkable from real
data. When $q\geqslant0$, both models reproduce quantitatively
similar $\tau(q)$ function of the return as real data, and the
$\tau(q)$ function for the volatility from the MF model deviates
from that of the real data while the modified MF model captures
excellently the multifractality in real data.

We have shown that our modified MF model is able to produce long
memory in the volatility while keeping the cubic law and
nonpersistence in the returns. The last but not least question is if
the long memory in the relative prices alone can reproduce the long
memory in the volatility when there is no memory in the order signs.
To address this question, we performed extensive simulations
following the MF model but with $H_s=0.5$ and $H_x=0.8$. We find
that the $H_v=0.78$, remaining unchanged when compared with the
modified model in which $H_s=0.75$ and $H_x=0.8$. Moreover, the
volatility is also distributed according to the cubic law. In
addition, we have $H_r=0.42$, indicating that the prices evolve in a
weak sub-diffusive behavior, which is nevertheless not far from the
diffusive regime with $H_r=0.5$. We note that some stocks do show
weak sub-diffusion effect
\cite{Bouchaud-Gefen-Potters-Wyart-2004-QF}.

\section{Concluding remarks}

In summary, we have improved the Mike-Farmer model for order-driven
markets by introducing long memory in the order aggressiveness,
which is an importnat stylized fact identified using the
ultra-high-frequency data of 23 liquid Chinese stocks traded on the
Shenzhen Stock Exchange in 2003. A detrended fluctuation analysis of
the relative prices $x$ unveils that $\overline{H}_x = 0.78 \pm
0.03$. The modified MF model is able to produce long memory in the
volatility with $\overline{H}_v = 0.79 \pm 0.02$, which is much
greater than $\overline{H}_v = 0.59 \pm 0.01$ obtained from the
original MF model. When we investigate the temporal correlation of
returns, we find that $\overline{H}_r = 0.53 \pm 0.01$, indicating
that the prices are diffusive. In addition, the cubic law for the
return distribution holds in the modified MF model. Our modified MF
model also enables us to distinguish the isolated memory effects of
order directions ($H_s$) and aggressiveness ($H_x$) on the
correlations in returns ($H_r$) and the volatility ($H_v$). We find
that $H_v$ is strongly dependent of $H_x$ and irrelevant to $H_s$.
In contrast, $H_r$ depends strongly on $H_s$ with little impact from
$H_x$. We confirmed that both the MF model and the modified MF model
are able to produce multifractality in the simulated prices.

The price formation process is fully determined by the dynamics of
order submission and order cancelation. Intuitively, the order
submission process has more important impact on the emergence of
long memory in the volatility. There are four factors in the order
submission process, the DFA scaling exponent $H_s$ of order signs,
the order size, the distribution $f(x)$ and the DFA scaling exponent
$H_x$ of relative prices. Our simulations show that the distribution
$f(x)$ might have impact on the return distribution
\cite{Gu-Zhou-2009-EPJB} but not the long memory in the volatility.
Therefore, we figure that the long memory of order aggressiveness is
a nontrivial main component of volatility clustering. To be more
rigorous, order size may be an alternative component of volatility
clustering. Indeed, order sizes are also long-term correlated
\cite{Lobato-Velasco-2000-JBES,Gopikrishnan-Plerou-Gabaix-Stanley-2000-PRE,Moyano-Souza-Queiros-2006-PA,Eisler-Kertesz-2006-EPJB,Eisler-Kertesz-2007-PA}
and there is well-established positive volume-volatility correlation
\cite{Karpoff-1987-JFQA}. However, the MF model and the modified MF
model do not include order size as an ingredient. This issue could
be addressed when a more realistic model is available, which is
beyond the scope of the current work.

\acknowledgments

We are grateful to the anonymous referees for their invaluable
suggestions. This work was partly supported by NSFC (70501011), Fok
Ying Tong Education Foundation (101086), NCET (07-0288), and
Shanghai Educational Development Foundation (2008SG29).

\bibliography{E:/Papers/Auxiliary/Bibliography}

\end{document}